\documentclass[twocolumn,floatfix,prl,showpacs,superscriptaddress]{revtex4}
\usepackage{graphicx,color,bm}

\begin{document}
\title{
Direct  determination of exchange parameters in Cs$_{2}$CuBr$_{4}$ and Cs$_{2}$CuCl$_{4}$: high-field ESR studies}

\author{S.A. Zvyagin}
\affiliation{Dresden High Magnetic Field Laboratory (HLD), Helmholtz-Zentrum
Dresden-Rossendorf, 01328 Dresden, Germany}

\author{D. Kamenskyi}
\altaffiliation[Present address: ]
{Radboud University Nijmegen, Institute for Molecules and Materials, High Field Magnet Laboratory, 6500 GL Nijmegen, The Netherlands}

\author{M. Ozerov}
\affiliation{Dresden High Magnetic Field Laboratory (HLD), Helmholtz-Zentrum
Dresden-Rossendorf, 01328 Dresden, Germany}

\author{J. Wosnitza}
\affiliation{Dresden High Magnetic Field Laboratory (HLD), Helmholtz-Zentrum
Dresden-Rossendorf, 01328 Dresden, Germany}
\affiliation{Instit\"{u}t fur Festk\"{o}rperphysik, TU Dresden, 01068 Dresden, Germany}
\author{ M. Ikeda}
\affiliation{KYOKUGEN, Osaka University, Toyonaka, Osaka 560-8531, Japan}

\author{T. Fujita}
\affiliation{KYOKUGEN, Osaka University, Toyonaka, Osaka 560-8531, Japan}

\author{M. Hagiwara}
\affiliation{KYOKUGEN, Osaka University, Toyonaka, Osaka 560-8531, Japan}

\author{A.I.~Smirnov}
\affiliation{P.L.~Kapitza Institute for Physical Problems, RAS, 119334 Moscow, Russia}

\author{T.A.~Soldatov}
\affiliation{Moscow Institute for Physics and Technology, 141700 Dolgoprudnyi, Russia}

\author{A.Ya.~Shapiro}
\affiliation{A.V.~Shubnikov Institute of Crystallography, RAS, 119333, Moscow, Russia}

\author{J. Krzystek}
\affiliation{National High Magnetic Field Laboratory, Florida
State University, Tallahassee, FL 32310, USA}

\author{R. Hu}
\altaffiliation[Present address: ]
{Rutgers Center for Emergent Materials and Department of Physics and Astronomy, Rutgers University, Piscataway, NJ 08854, USA }
\affiliation{Condensed Matter Physics and Materials Science Department, Brookhaven
National Laboratory, Upton, NY 11973, USA}

\author{H. Ryu}
\affiliation{Condensed Matter Physics and Materials Science Department, Brookhaven
National Laboratory, Upton,
NY 11973, USA}
\affiliation{Department of Physics and Astronomy, Stony Brook University, Stony Brook,
New York 11794-3800, USA}

\author{C. Petrovic}
\affiliation{Condensed Matter Physics and Materials Science Department, Brookhaven
National Laboratory, Upton, NY 11973, USA}
\affiliation{Department of Physics and Astronomy, Stony Brook University, Stony Brook,
New York 11794-3800, USA}

\author{M.E. Zhitomirsky}
\affiliation{Service de Physique Statistique, Magn\'etisme et Supraconductivit\'e,
UMR-E9001 CEA-INAC/UJF,  38054 Grenoble Cedex 9, France}

\begin{abstract}
Spin-1/2  Heisenberg antiferromagnets Cs$_2$CuCl$_4$ and Cs$_2$CuBr$_4$ with distorted triangular-lattice
structures are studied by means of  electron spin resonance spectroscopy in magnetic fields up to the saturation field and above.   In the  magnetically saturated phase,   quantum fluctuations are
fully  suppressed, and the spin dynamics is defined by ordinary magnons. This allows us
to accurately describe the magnetic excitation spectra in both materials and, using the harmonic spin-wave theory,
to  determine their exchange parameters. The viability of the proposed
method was proven by applying it to Cs$_2$CuCl$_4$, yielding $J/k_B=4.7(2)$~K,
$J'/k_B=1.42(7)$~K  [$J'/J\simeq 0.30$]  and  revealing good agreement with inelastic neutron-scattering
results. For the isostructural   Cs$_2$CuBr$_4$,   we obtain  $J/k_B=14.9(7)$ K,  $J'/k_B=6.1(3)$ K,
[$J'/J\simeq 0.41$], providing  exact and conclusive information on  the exchange couplings in this  frustrated
spin system.
\end{abstract}

\maketitle

A spin-1/2 Heisenberg antiferromagnet (AF) on a triangular lattice is the paradigmatic model
in quantum magnetism, which was intensively studied since  Anderson's conjecture of
the resonating-valence-bond ground state  \cite{Anderson}.   In  spite of
numerous theoretical  studies (which predict  a rich variety of grounds states, ranging from a gapless spin liquid to  N\'{e}el order),
many important details of the phase diagram of
triangular-lattice AFs remain controversial or even missing (see, i.e., \cite{Weng,Schmied,Heid,Starykh,Gham,Weich}).

In order to test the theory experimentally, a precise information on the spin-Hamiltonian parameters for the  materials of interest is highly demanded.
The presence of quantum fluctuations  makes the accurate  description of such systems (first of all, the extraction of the spin
Hamiltonian parameters)  extremely challenging.
One solution to solve this problem   is
to suppress quantum fluctuations by strong-enough magnetic fields. The system is then in  the
spin-polarized, magnetically saturated phase. The  excitation spectrum above the saturation field,  $H_{sat}$, is determined by ordinary magnons, which can be described
quantitatively by a simple harmonic spin-wave theory.

Studying the magnon dispersion  in quantum magnets above   $H_{sat}$  by means of inelastic neutron-scattering provides
the most straightforward opportunity to extract  parameters of the spin Hamiltonian.
This method   has been used, for instance, to determine the exchange coupling parameters in the triangular-lattice AF Cs$_2$CuCl$_4$
\cite{Coldea_INS}. Experiments revealed up to $65\%$ difference between the parameters estimated
at $H=0$ (using the harmonic approximation) and actual values (extracted from measurements at $H>H_{sat}$),
stressing the great importance of  high-field  experiments. Unfortunately, the applicability of this technique  is limited
to magnetic fields (of about 15 T) currently available for neutron-scattering experiments.

Electron spin resonance (ESR) offers another means to measure  the spin
Hamiltonian parameters,  directly and with  high precision.  Similar to the case  of neutron scattering, the distinct
advantage of the high-field ESR  is the availability of $exact$
theoretical spin-wave expressions for the magnetically saturated phase. For instance, measurements of  ESR
spectra in the spin-1 material  $\rm NiCl_2$-$\rm 4SC(NH_2)_2$ (known as DTN)  above the saturation field,
$H_{sat}=12.6$ T, allowed to determine the
bare single-ion anisotropy and, based on that, to accurately  describe
the temperature-field phase diagram \cite{Zvyagin_DTN}.

In this Letter, we report on a new approach, which combines high-field ESR  as a tool
to probe the magnon-excitation spectrum  above $H_{sat}$ and its
classical linear spin-wave description, allowing us to accurately determine exchange coupling parameters in a spin-1/2
Heisenberg triangular-lattice AF. This approach is based on the observation of  ESR modes of a new type,
which becomes possible due to the low-enough crystal symmetry of the studied materials.
First, we proved the viability of the proposed technique  by applying it to Cs$_2$CuCl$_4$.
Good agreement between the neutron-scattering \cite{Coldea_INS} and ESR results was obtained.
Then, this procedure was employed for the determination of the exchange parameters in the isostructural compound Cs$_2$CuBr$_4$, providing
the direct answer to the long-standing problem of the spin Hamiltonian parameters of this frustrated
compound.

\begin{figure}[t]
\begin{center}
\vspace{-1.5cm}
\includegraphics[width=0.5\textwidth]{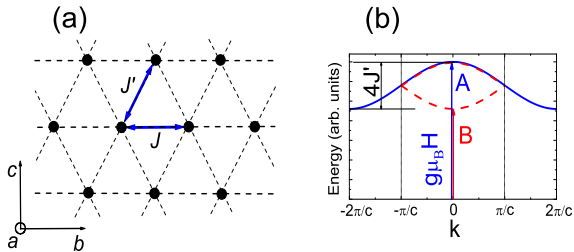}
\vspace{-2.5cm}
\caption{\label{fig:str+disp} (Color online) (a) Schematic picture of
exchange paths in the $bc$ plane of CCC and CCB. (b) Dispersion of magnon excitations for a spin-1/2 Heisenberg AF with  triangular  lattice  in the saturated phase for an arbitrary magnetic field.  Solid blue line
is the dispersion of magnon excitations in the exchange approximation  (Eq. \ref{ccc_disp}).   The magnon dispersion within the folded Brillouin zone is shown by the dashed red line.  Arrows A and
B correspond to the observed ESR transitions.}
\end{center}
\end{figure}

In spite of the recent progress in synthesizing new spin-1/2 triangular-lattice materials
(see \cite{Yamashita_1,Yamashita_2,Kanoda,Powel,BCSO} and reference herein) the two compounds,
Cs$_2$CuCl$_4$ and Cs$_2$CuBr$_4$ (hereafter called CCC and CCB),
remain among the most prominent representatives of such kind of frustrated magnets.
The Cu$^{2+}$  ions in CCC and CCB form a distorted triangular lattice
and can be described by the  exchange Hamiltonian
\begin{equation}
\mathcal{H} = J\sum_{\langle i,j \rangle} {\bf S}_i\cdot{\bf S}_j +
J' \sum_{\langle i,j' \rangle}  {\bf S}_i\cdot{\bf S}_{j'} \ , 
\label{Ham}
\end{equation}
where  ${\bf S}_i$, ${\bf S}_j$,  and ${\bf S}_{j'}$ are spin-1/2  operators at sites $i$, $j$, and $j'$, respectively;   $J$ is the interaction constant along the $b$ direction;  $J'$ is the zig-zag interchain
coupling [Fig.~\ref{fig:str+disp} (a)]. The orthorhombic crystal structure of CCC corresponds to the space group $Pnma$ with
the room-temperature lattice parameters $a = 9.769$~{\AA},  $b = 7.607$~{\AA}, and
$c = 12.381$~{\AA} \cite{CCC_CP}. At $T_N=0.62$ K, CCC  undergoes a transition into  helical incommensurate
(IC) long-range-ordered state
\cite{Coldea_TN}. CCC is in the saturated
phase above the critical fields $H_{sat}=8.44$, 8.89 and 8~T applied along the $a$, $b$, and $c$ axis,
respectively \cite{Tokiwa_PT}. The exchange interactions were estimated from the mentioned above inelastic neutron-scattering experiments in the saturated phase \cite{Coldea_INS}, yielding  $J/k_B=4.34(6)$ K and  $J'/k_B=1.48(6)$ K  [$J'/J=0.34(2)$].

Similar to CCC, the  isostructural compound CCB realizes a distorted triangular lattice with
the room-temperature lattice  parameters $a = 10.195$~{\AA}, $b = 7.965$~{\AA}, and
$c = 12.936$~{\AA} \cite{Morosin}. At $T_N = 1.4$ K, CCB undergoes a transition into  helical
IC long-range-ordered state
\cite{Ono_ENS,Ono_PRB}. CCB is in the  saturated phase above  $H_{sat}=30.71$,
30.81, and 28.75~T applied along the $a$, $b$, and $c$ axis, respectively \cite{Ono_INS}.
Within a classical spin model, the ratio
$J'/J=0.467$ was estimated \cite{Ono_ENS,Ono_PRB,Ono_INS}. On the other hand, results of  density-functional
calculations suggest  $J'/J\sim 0.5-0.65$ \cite{Valenti}, while the ratio $J'/J=0.74$  was obtained \cite{Tsujii}  by comparison of the zero-field IC wavenumber in the ordered phase  with results of the series expansion method \cite{Weihong}. 

Single crystals of CCC (CCB) were synthesized by slow evaporation of aqueous solutions of
CsCl and CuCl$_2$  (CsBr and CuBr$_2$). Samples of CCC were from the same batch as in Ref. \cite{Povarov,Smirnov}. Experiments  were performed using ESR spectrometers operated in combination with superconducting (KYOKUGEN, HLD, Kapitza Institute), 25 T resistive (NHMFL \cite{spectrometer}), and 50 T pulse-field (KYOKUGEN, HLD) magnets.  The spectrometer at the Kapitza Institute with a $^3$He insert  and 12 T magnet was used
for taking spectra  down to 0.45 K. Backward wave oscillators, VDI generators (product of
Virginia Diodes Inc.), and CO$_2$-pumped molecular laser (product of Edinburgh Instruments Ltd.) were used as sources of mm- and submm-wavelength radiation. In our experiments, the magnetic field was applied along the crystallographic $b$ axis.  2,2-diphenyl-1-picrylhydrazyl (known as DPPH) was employed as a standard marker for the accurate calibration of the magnetic field.

\begin{figure}[t]
\begin{center}
\vspace{0cm}
\includegraphics[width=0.45\textwidth]{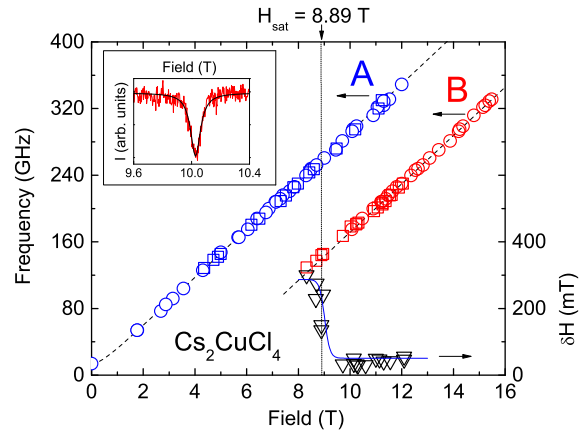}
\vspace{-0.5cm}
\caption{\label{fig:ccc_ffd} (Color online)
Frequency-field diagram of the ESR excitations in CCC
 measured at 0.5 K  (squares) and 1.5 K (circles). Dashed lines correspond to fit results (see text for details). The linewidth (half width at half maximum) of mode B vs field is shown by triangles;  the solid line  is a guide for
the eye. Inset shows an example of ESR spectrum (mode B) taken at  178.3 GHz ($T=0.53$ K);   the solid line corresponds to a Lorentzian fit. }
\end{center}
\end{figure}

\begin{figure}[t]
\begin{center}
\vspace{-0.5cm}
\includegraphics[width=0.5\textwidth]{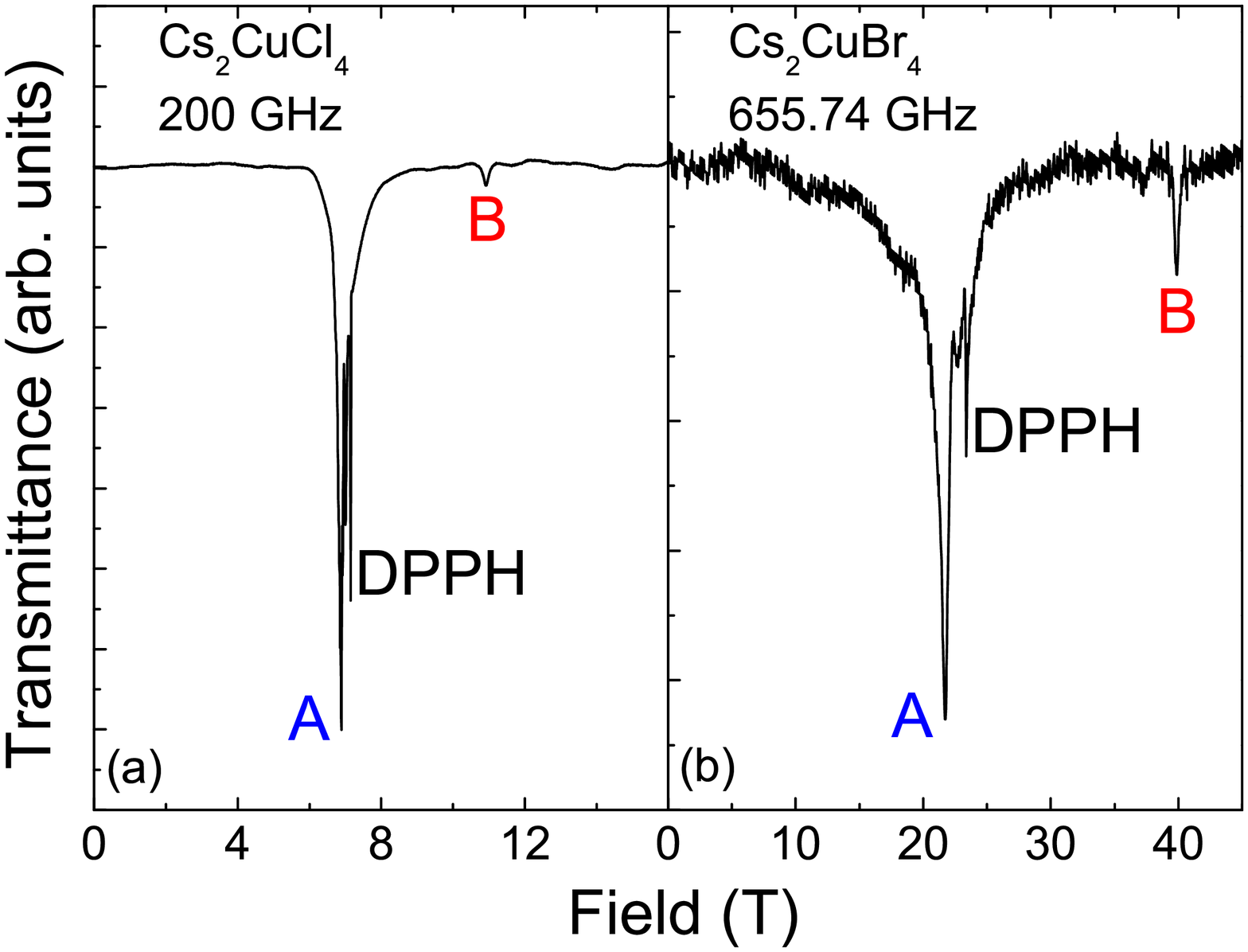}
\vspace{-0.8cm}
\caption{\label{fig:ccc_ccb_spectra} (Color online) ESR spectra of CCC (a) and CCB (b),
 taken at frequencies 200 and 655.74 GHz, respectively ($T=1.5$ K). DPPH is employed as a standard marker.
}
\end{center}
\end{figure}

The frequency-field diagrams of  the ESR absorption in CCC at  0.5 and 1.5 K are  shown in
Fig.~\ref{fig:ccc_ffd} by squares and circles, respectively. Two resonance modes of different intensity were observed (Fig.~\ref{fig:ccc_ccb_spectra}, a). The most intensive mode, the mode A, can be
described using the equation $\hbar\omega_A = \sqrt{(g_b\mu_B H)^{2}+\Delta_A^{2}}$,
where $\hbar$ is the Planck constant, $\omega$ is the excitation frequency, $\mu_B$ is the Bohr
magneton, $\Delta_A/(2\pi\hbar)=14$~GHz and  $g_b=2.08(2)$ \cite{Povarov}. Above $H_{sat}$, the mode A
corresponds to the collective excitation of spins with the frequency
$\omega_A \approx g_b\mu_B H/\hbar$ and can be interpreted  as uniform \textbf{k} = 0  precession of
spins around  the field direction. The much weaker mode  B appears at $H\gtrsim H_{sat}$. The  frequency of this mode can be described empirically using the equation
$\hbar \omega_B =  g_b\mu_B H - \Delta_B$ with the same $g$-factor, $g_b=2.08$, and $\Delta_B/(2\pi\hbar)=119.0(3)$~GHz. The ESR line undergoes a significant broadening approaching $H_{sat}$ from the high-field end (Fig.~\ref{fig:ccc_ffd}, triangles), becoming undetectable  below 8 T. An example of ESR spectrum (mode B),  taken at  178.3 GHz ($T=0.53$ K) is shown  in the   inset of  Fig.~\ref{fig:ccc_ffd};  the solid line corresponds to a Lorentzian fit.

The emergence of two ESR modes in the magnetically  saturated state
signifies a lower crystal symmetry compared to the one
assumed  in the simple spin model (\ref{Ham}).
This is not entirely surprising since the unit cell
of CCC is made up of four
inequivalent Cu$^{2+}$ ions: two on the adjacent $b$-chains in the $bc$ plane and
two on the adjacent layer shifted along the $b$ and $c$  axes  by a half of lattice
constant  \cite{Coldea_TN}.
One  single copper layer is described by the exchange Hamiltonian
(\ref{Ham}).  On the other hand, the crystal symmetry of CCC allows the Dzyaloshinskii-Moriya (DM) interaction for all nearest-neighbor
spin  bonds in a single copper layer:
\begin{equation}
\hat{\mathcal{H}}_{DM} = \sum_i \sum_{n=1}^3 {\bf D}_n \cdot \bigl[{\bf S}_i \times
{\bf S}_{i+\bm{\delta}_n} \bigr] \ ,
\label{HDM}
\end{equation}
where the lattice vectors $\bm{\delta}_n$  are chosen as
$\bm{\delta}_1 = (0,b,0)$, $\bm{\delta}_{2,3} = (0,\pm b/2,c/2)$.
The DM vectors compatible with the space group of the crystal
are given by
\begin{eqnarray}
&& {\bf D}_1  =  ( D_a,0,(-1)^{i_c} D_c) \ ,
\label{DM} \\
&& {\bf D}_{2,3} =  (\pm D'_a,(-1)^{i_c} D'_b, \pm(-1)^{i_c} D'_c) \ ,
\nonumber
\end{eqnarray}
where $i_c$ is the chain index in the $c$-direction (see \cite{Starykh} for further details on the DM interactions in CCC).
So far, experiments on CCC gave estimates for  three
DM parameters: $D_a'/J \sim 5$\% \cite{Coldea_INS} and
$D_{a,c}/J \sim 10$\% \cite{Povarov,Fayzullin}.

The reduced translational symmetry
of the copper layers in CCC (CCB)
revealed by the staggered DM vectors (\ref{DM}) leads to the folding of
the Brillouin zone of a simple triangular Bravais lattice.
As a result, the ESR transitions are allowed not only for
${\bf k}=0$ (mode A) but also for $k_c=2\pi/c$ (exchange mode B).
A detailed analysis of the excitation spectrum for the spin Hamiltonian
given by the sum of (\ref{Ham}) and (\ref{HDM}) is presented in the Supplemental
Material \cite{SM}. Here, we resort to a simpler line of arguments valid
in the case of small DM interaction. We just neglect the effect of the DM terms
(\ref{HDM}) on the magnon energy. Then, the dispersion of the magnetic excitations
for a spin-1/2 AF (\ref{Ham}) in the saturated phase is described by

\begin{equation}
\textstyle
\label{ccc_disp}
\hbar\omega_{\bf k} = g\mu_B H  + J \cos(k_b b) + 2J' \cos(\frac{1}{2}k_b b)
\cos(\frac{1}{2}k_c c)  - J_0 \ ,
\end{equation}
where $J_0 =J + 2J'$. The difference between the excitation energies of the modes A and B [Fig.~\ref{fig:str+disp} (b)]
is equal to
\begin{equation}
\hbar \Delta \omega = 4J' \ .
\label{Dw}
\end{equation}
For ${\bf H}\parallel b$, the above approximate expression can be compared to the
exact result \cite{SM}:
\begin{equation}
\hbar \Delta \omega = 4\sqrt{(J')^2 + (D_b')^2} \ .
\end{equation}
For this, as well as for other field orientations, the  correction
from a finite value of the DM interaction is
of the order of $(D_b'/J')^2$ and does not exceed 1--2\%.  However,
a finite value of $D_b'$ is essential for the  observation of
mode B: the intensity ratio of the two resonance lines
scales as $(D_b'/J')^2$, so that the mode B would not be seen
for $D_b'=0$. Hence, measurements of the ESR spectra  in the saturated phase provide
a direct and accurate estimate of  $J'$.

Knowing $J'$, we now can determine $J$ from the  saturation field
using the expression
\begin{equation}
g\mu_B H_{sat} = 2J(1+J'/2J)^2
\label{ccc_sat}
\end{equation}
obtained for the exchange model (\ref{Ham}).
The correction to Eq.~(\ref{ccc_sat}) taking the DM interactions into account
can be assessed using the expression
obtained for ${\bf H}\parallel b$ \cite{SM}:
\begin{equation}
g\mu_B H_{sat}^b = 2(J+J') + (J'^2+D_b'^2)/2J  \ .
\end{equation}
Even for $D_b'\sim 0.1$--$0.2J$, the effect on $H_{sat}$  for CCC(CCB)  can be safely neglected.
Thus, using Eqs.~(\ref{ccc_disp}) and (\ref{ccc_sat}),
$g_b=2.08(2)$,  and $H_{sat}=8.89(2)$ T, the exchange coupling parameters
for CCC are obtained as
$J/k_B=4.7(2)$ K and $J'/k_B=1.42(7)$ K [$J'/J=0.30(3)$] \cite{acc}.
The latter value is in good agreement with the estimate $J'/J=0.34(2)$ from
the neutron-scattering experiments \cite{Coldea_INS}.

Let us also note  that in CCC (CCB) the Brillouin-zone folding  occurs also in the $a$ direction perpendicular
to the copper layers. By a similar line of arguments this folding yields a further splitting of each mode A and B
by $\delta\omega'' \sim J''$, where $J''$ is the interlayer  exchange coupling. Since $J''=0.2$~K \cite{Coldea_INS},
it is rather difficult to observe  such a splitting even in ESR experiments.

\begin{figure}[t]
\begin{center}
\vspace{-0cm}
\includegraphics[width=0.5\textwidth]{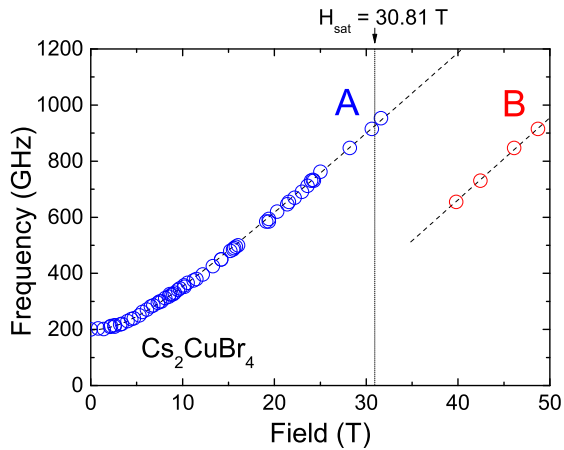}
\vspace{-0.8cm}
\caption{\label{fig:ccb_ffd} (Color online) Frequency-field diagram of ESR excitations in CCB
($T=1.5$ K).   Dash lines correspond to fit results (see the text for details). }
\end{center}
\end{figure}

Once the viability of the proposed approach is verified, we apply it now to CCB.
Compared to CCC, $H_{sat}$ in CCB is more than 3 times larger, that implies
the necessity of ESR measurements in magnetic field above 30 T.  Similar to CCC, two ESR modes
have been observed at $H>H_{sat}$ (Fig.\ \ref{fig:ccc_ccb_spectra}, b). The frequency-field
diagram of ESR excitations in CCB obtained at $T=1.5$ K is shown in Fig.\ \ref{fig:ccb_ffd}. The mode A can be described using the equation
$\hbar\omega_A = \sqrt{(g\mu_B H)^{2}+\Delta_A^{2}}$, where  $g_b=2.09(2)$  and $\Delta_A/(2\pi\hbar) = 198$ GHz.  The exchange mode B was observed only above $H_{sat}$. The  mode B can be
described by the equation $\hbar \omega_B = g_b\mu_B H - \Delta_B$, where $\Delta_B/(2\pi\hbar) = 507.6$ GHz.
Using Eqs.~(\ref{ccc_disp}) and (\ref{ccc_sat}), the exchange coupling parameters for CCB are
obtained:  $J/k_B=14.9(7)$ K and  $J'/k_B=6.1(3)$ K [$J'/J=0.41(4)$] \cite{acc}.

As mentioned, our results  for  CCC are in a good agreement with those obtained earlier using inelastic neutron-scattering experiments \cite{Coldea_INS}. On the other hand, in the case of CCB a relatively big difference between previously suggested value, $J'/J=0.74$ \cite{Ono_INS},  and our result, $J'/J=0.41(4)$, is  observed. This difference is of  crucial importance for understanding  of unusual magnetic properties of CCB.  For instance,  CCB is a rare example of  a spin-1/2 triangular-lattice Heisenberg AF, which exhibits a 1/3 magnetization plateau \cite{Ono_ENS,Ono_INS,Ono_PRB}. Numerical diagonalization
calculations of a finite-size spin-1/2 Heisenberg AF predicts that the geometric frustration should be sufficiently strong to stabilize the so-called ``up-up-down'' (UUD) phase, resulting in the emergence of the 1/3 magnetization plateau,  in the range $0.7\lesssim J'/J \lesssim 1.3$ \cite{Miyahara}. On the other hand, density matrix renormalization  group calculations predict the 1/3 magnetization plateau even for infinitesimally small $J'/J$ ratio \cite{Chen}. Our results suggest  that  the field-induced UUD phase in spin-1/2 triangle-lattice Heisenberg AF can be realized for the $J'/J$ ratio, which is much smaller than predicted in Ref. \cite{Miyahara}. The obtained spin-Hamiltonian parameters can  be of particular importance for a quantitative  description of the cascade of  field-induced phase transitions observed recently in CCB \cite{Fortune}.

In conclusion, the excitation spectra of Cs$_{2}$CuCl$_{4}$ and Cs$_{2}$CuBr$_{4}$ have been probed in  magnetic fields up to $H_{sat}$ and above.
Based on  the classical linear spin-wave description of the magnon excitation spectrum  and  high-field magnetization data, the exchange coupling parameters for both compounds were determined. The obtained accurate knowledge is of eminent  importance for the understanding of the complex  phase diagram  of spin-1/2 triangular-lattice Heisenberg AFs. The proposed approach can be used for accurate estimation of exchange parameters of a growing family of spin-1/2 triangular-lattice AFs, including  organic compounds  (see \cite{Yamashita_1,Yamashita_2,Kanoda,Powel} and references herein),  those investigations via conventional neutron-scattering techniques is rather challenging. The employment of very high magnetic fields  (up to ca 70 T  \cite{FEL,Zvyagn_CuPM} and above \cite{Nojiri,Kindo,Boriskov}, currently available for pulsed-field magneto-spectroscopy) as well as the rapid progress in the THz techniques makes the proposed method of crucial importance  for investigating spin systems with large $J/k_B$.
The  approach has  a broader impact and can be potentially used for $any$ quantum magnet  with reduced (e.g., by the staggered DM interaction) translational symmetry, resulting, as predicted,  in emergence of a new exchange mode above $H_{sat}$.

This work was supported in part  by the DFG. We acknowledge the support of the HLD at HZDR,
member of the European Magnetic Field Laboratory. S.A.Z. appreciates the support of the Visiting Professor
Program at KYOKUGEN in Osaka University. Work at Brookhaven was supported by
the U.S. DOE under Contract No. DE-AC02-98CH10886. C.P. acknowledges the support by the A. von Humboldt Foundation.
 Work at the Kapitza Institute
is supported by Russian foundation for basic research, grant No. 12-02-00557.  A portion of this work was
 performed at the NHMFL, which is supported by NSF Cooperative
  Agreement No. DMR-1157490, by the State of Florida, and by the DOE.  The authors would like to thank V.N. Glazkov, A.K. Kolezhuk, V.I. Marchenko,  S.S. Sosin,  and O.A. Starykh for discussions, and S. Miyasaka for the help in orienting the CCB samples.

\onecolumngrid

\section{SUPPLEMENTAL MATERIAL}
\section{2D model for $\rm Cs_2CuCl_4$}

\begin{figure}[b]
\centerline{
\includegraphics[width=0.4\columnwidth]{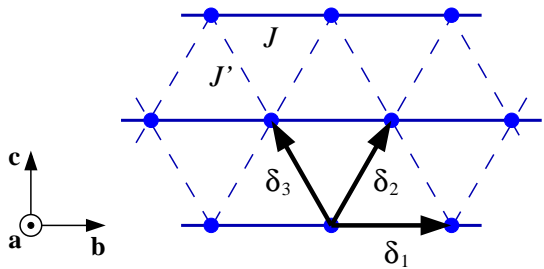}}

\caption{The orthorhombically distorted triangular lattice.
}
\label{fig:model}
\end{figure}

A single plane of copper spins in $\rm Cs_2CuCl_4$ is described by
the following spin Hamiltonian \cite{Coldea02,Starykh10}:
\begin{eqnarray}
\hat{\cal H} & = & \sum_i \Bigl\{ J {\bf S}_i \cdot {\bf S}_{i+\bm{\delta}_1}
+ J' \bigl({\bf S}_i \cdot {\bf S}_{i+\bm{\delta}_2} +
{\bf S}_i \cdot {\bf S}_{i+\bm{\delta}_3}\bigr)
+ {\bf D}_1 \cdot \bigl[{\bf S}_i \times {\bf S}_{i+\bm{\delta}_1}\bigr]
\nonumber \\
& & \phantom{\sum_i  J {\bf S}_i \cdot  }
+ {\bf D}_2 \cdot \bigl[{\bf S}_i \times {\bf S}_{i+\bm{\delta}_2}\bigr]
+ {\bf D}_3 \cdot \bigl[{\bf S}_i \times {\bf S}_{i+\bm{\delta}_3}\bigr]
- g\mu_B\,{\bf H}\cdot {\bf S}_i\Bigr\} \,,
\label{Hspin}
\end{eqnarray}
where the in-plane nearest-neighbor vectors
are defined as $\bm{\delta}_1 = (0,b,0)$ and
$\bm{\delta}_{2,3} = (0,\pm b/2,c/2)$, see Fig.~\ref{fig:model}.
From now on we shall use the short-hand notation $g\mu_BH\to H$.
The two exchange constants $J$ and $J'$ correspond to  the chain and zigzag bonds,
respectively. The Dzyaloshinskii-Moriya (DM) vector on the chain bonds is
\begin{equation}
{\bf D}_1 = \bigl(D_a, 0, (-1)^{i_c} D_c)\,,
\end{equation}
where $i_c$ is the chain index in the $c$  direction.
On the interchain zigzag bonds one  has by symmetry
\cite{Starykh10}
\begin{equation}
{\bf D}_{2,3} = \bigl(\pm D_a',(-1)^{i_c} D_b', \pm (-1)^{i_c} D_c')\ .
\end{equation}

The Hamiltonian (\ref{Hspin}) is translationally invariant in the
direction of spin chains ($b$-axis), whereas the alternation of the
components of the Dzyaloshinskii-Moriya vectors between the chains leads
to a unit cell with two copper atoms.

\section{High-field two-sublattice structure}

In antiferromagnets with a uniform arrangement of DM-vectors all spins become
parallel to each other at $H>H_s$, though this direction generally deviates
from the field direction unless $\bf H\parallel {\bf D}$.
The alternation of the DM vectors in the spin Hamiltonian (\ref{Hspin}) leads to
the presence of two distinct sublattices even in high magnetic fields.
It is important that the two sublattices correspond to
adjacent  chains, whereas the translational symmetry along the chains remains
unbroken.

Let us check which components of the DM vectors in (\ref{Hspin}) are  responsible
for the above effect and, therefore,  play a role in the uniform magnetic resonance.
The DM interaction on the chain bonds can be rewritten as
\begin{equation}
\hat{\cal H}_{DM}^{(1)}  =  \frac{1}{2} \sum_i   {\bf D}_1\cdot \bigl[ {\bf S}_i \times
( {\bf S}_{i+\bm{\delta}_1} - {\bf S}_{i-\bm{\delta}_1}) \bigr] \ .
\label{DM1}
\end{equation}
Its expectation value  vanishes for a uniform state along the chains.
Hence,  the term (\ref{DM1}) may contribute to the ESR frequencies only
beyond the harmonic approximation and can be safely neglected at high fields.
On the zigzag interchain bonds
one has to consider separately different components. The contribution
from the $a$ component is
\begin{equation}
\hat{\cal H}_{DM}^{(2)}  =  \frac{1}{2}\, D_a' \sum_i   \hat{\bf a}\cdot\bigl[  {\bf S}_i \times
({\bf S}_{i+\bm{\delta}_2} - {\bf S}_{i-\bm{\delta}_2} - {\bf S}_{i+\bm{\delta}_3}
+ {\bf S}_{i-\bm{\delta}_3} )\bigr] \ .
\end{equation}
It again vanishes for a translationally invariant state along the chains.
The contribution from the $b$ component is
\begin{equation}
\hat{\cal H}_{DM}^{(3)}  =  \frac{1}{2}\, D_b' \sum_i   \hat{\bf b}\cdot\bigl[  {\bf S}_i \times
({\bf S}_{i+\bm{\delta}_2} + {\bf S}_{i-\bm{\delta}_2} + {\bf S}_{i+\bm{\delta}_3}
+ {\bf S}_{i-\bm{\delta}_3} )\bigr] \ .
\label{HDMb}
\end{equation}
This contribution does not vanish and generates a staggered field along the $b$ direction
responsible for the two sublattice structure, which survives in strong magnetic field.
Finally,  the contribution from the $c$ component is
\begin{equation}
\hat{\cal H}_{DM}^{(4)}  =  \frac{1}{2}\, D_c' \sum_i   \hat{\bf c}\cdot\bigl[  {\bf S}_i \times
({\bf S}_{i+\bm{\delta}_2} + {\bf S}_{i-\bm{\delta}_2} - {\bf S}_{i+\bm{\delta}_3}
- {\bf S}_{i-\bm{\delta}_3} )\bigr] \ ,
\end{equation}
which again does not change the ground state and produces only cubic terms in
the bosonic representation of the uniform resonance modes.

Thus, the ESR spectrum in the high-field state can be studied with a simplified
two-sublattice representation of the full spin Hamiltonian (\ref{Hspin}):
\begin{equation}
\hat{\cal H}_{2s}  =  4J'\,{\bf S}_1 \cdot {\bf S}_2 + 4 D_b' \;\hat{\bf x}\cdot\bigl[{\bf S}_1 \times
{\bf S}_2 \bigr]  - {\bf H}\cdot ({\bf S}_1 + {\bf S}_2) \ .
\label{H2spin}
\end{equation}
Let us emphasize again that the expression (\ref{H2spin}) is applicable only for studying ESR
modes with ${\bf k} =0$ in the folded Brillouin zone.
The saturation field $H_s$, which corresponds to magnon condensation at
an incommensurate value of $k_b$ \cite{Coldea02},  must be obtained from the full
spin Hamiltonian (\ref{Hspin}), see Sec.~IV.

\section{High-field magnetic resonance}

We now investigate modes of the ESR resonance for the two-sublattice antiferromagnet
with the Dzyaloshinskii-Moriya interaction (\ref{H2spin}) in the high-field
paramagnetic state. In the following, the vector components $x$, $y$, and $z$
refer to the rotated coordinate frame with the $z$ axis being parallel to the field
direction.

\subsection{Collinear geometry}

Let us begin with the collinear geometry ${\bf H}\parallel{\bf D}$. In this case
$S^z_{\rm tot}$ is a good quantum number and for $H>H_s$ all spins become parallel
to $\bf H$. Using the standard Holstein-Primakoff transformation and keeping only
harmonic terms we obtain for Eq.~(\ref{H2spin}):
\begin{eqnarray}
\hat{\cal H}_2 & = &4J'S\, ( a^\dagger_1 a_2 + a^\dagger_2 a_1 - a^\dagger_1 a_1 - a^\dagger_2 a_2)
+ 4D_b'Si\, (a^\dagger_1 a_2 - a^\dagger_2 a_1) + H \,(a^\dagger_1 a_1 + a^\dagger_2 a_2) \,.
\label{H22}
\end{eqnarray}
Straightforward diagonalization yields two resonance modes
\begin{equation}
\varepsilon_{1,2} =  H - 4J'S \pm 4S \sqrt{J'^2+D_b'^2} \,,
\label{E22}
\end{equation}
such that the splitting between the two resonance modes at any given $H>H_s$ is
\begin{equation}
\varepsilon_{1} -\varepsilon_{2} =  8S \sqrt{J'^2+D_b'^2} \,.
\end{equation}
For $S=1/2$ and weak DM interactions, the mode splitting is approximately given by
$$
\Delta \varepsilon \approx 4J' [1 + D_b'^2/(2J'^2)] \,.
$$

By explicitly writing the diagonalization transformation for Eq.~(\ref{H22})
we can also determine the intensity ratio for the two modes. First, one performs
the transformation to symmetric/antisymmetric bosons:
\begin{equation}
\textstyle
b_1 = \frac{1}{\sqrt{2}} (a_1+a_2) \ , \quad
b_2 = \frac{1}{\sqrt{2}} (a_1-a_2) \,,
\end{equation}
which yields
\begin{eqnarray}
\hat{\cal H}_2 & = & H b^\dagger_1 b_1 + (H-8J'S) b^\dagger_2 b_2
+ 4D_b'Si\, (b^\dagger_2 b_1 - b^\dagger_1 b_2)\ .
\end{eqnarray}
The RF field with $q=0$ couples to the symmetric $b_1$ boson. Hence, the ESR absorption
is given by the imaginary part of the transverse uniform susceptibility
\begin{equation}
\chi''(q=0,\omega) = - \frac{1}{\pi}\, \textrm{Im}\{G_1(\omega)\} \ , \qquad
G_1(t) = -i \langle T b_1(t) b_1^\dagger\rangle \ .
\end{equation}
Subsequent transformation to the eigenmodes (\ref{E22}) is performed with
the generalized operator rotation
\begin{equation}
b_1 = uc_1 + ivc_2 \ , \quad
b_2 = uc_2 + ivc_1 \ , \qquad  \textrm{with} \quad u^2+v^2=1 \ .
\end{equation}
Choosing
\begin{equation}
2uv = \frac{D_b'}{\sqrt{J'^2+D_b'^2}} \,,
\end{equation}
we arrive at the spectrum (\ref{E22}). The susceptibility on the other hand is given
by a sum of two delta-functions:
\begin{equation}
\chi''(\omega) = u^2 \delta(\omega-\varepsilon_1) + v^2  \delta(\omega-\varepsilon_2)\ .
\end{equation}
Since $v\approx D_b'/(2J')$, the intensity ratio for the two modes scales as
$$
I_2/I_1 \sim (D_b'/2J')^2 \ .
$$

\subsection{Orthogonal geometry}

Now consider the orthogonal geometry $\bf H\perp D$. In this case the two-sublattice
structure,
\begin{equation}
{\bf S}_1 = (-\cos\theta, 0,\sin\theta) \,, \qquad
{\bf S}_2 = (-\cos\theta, 0,\sin\theta) \,,
\end{equation}
with $\theta\to \pi/2$ survives to arbitrary strong magnetic fields.
The classical energy is
\begin{equation}
E_c = -4J'S^2\cos2\theta  - 4D_b'S^2\sin2\theta - 2HS \sin\theta \ .
\end{equation}
Its minimum is achieved for
\begin{equation}
\sin\theta - \frac{D_b'}{2J'} \frac{\cos 2\theta}{\cos\theta} = \frac{H}{8J'S}\equiv h \ .
\end{equation}
For $h\gg 1$, one finds $\theta=\pi/2 -\alpha$ with $\alpha \approx D_b'/[2J'(h-1)]$.

Transforming for each sublattice  to the local twisted frame we obtain with
the required accuracy
\begin{equation}
\hat{\cal H}_{2s}  =  4J'\bigl[ S_1^yS_2^y - \cos2\theta (S_1^xS_2^x+S_1^zS_2^z)\bigr]
- 4D_b' \sin2\theta ( S_1^xS_2^x+S_1^zS_2^z) -H\sin\theta (S_1^z+S_2^z) \ ,
\label{H2perp}
\end{equation}
which yields the harmonic magnon Hamiltonian
\begin{eqnarray}
\hat{\cal H}_2 & = & A  ( a^\dagger_1 a_1 + a^\dagger_2 a_2)
+ C  ( a^\dagger_1 a_2 + a^\dagger_2 a_1) - B  ( a_1 a_2 + a^\dagger_2  a^\dagger_1) \,,
\label{H2pp}
\\
&&
A  =  4J'S + 4D_b'S \tan\theta \,,\ \
B = 4S\cos\theta (J'\cos\theta + D_b'\sin\theta) \,,\ \
C = 4S\sin\theta (J'\sin\theta - D_b'\cos\theta) \,.
\nonumber
\end{eqnarray}
The diagonalization of (\ref{H2pp}) is somewhat more involved but is still straightforward.
It yields two resonance modes
\begin{equation}
\varepsilon_{1,2} = \sqrt{(A\pm C)^2 - B^2} \ ,
\end{equation}
or, denoting $d = D_b'/J'$,
\begin{equation}
\varepsilon_{1} = 4J'S\sqrt{(2\sin^2\theta - d \tan\theta\cos2\theta)(2+ d\tan\theta)} \ , \quad
\varepsilon_{2} = 4J'S\sqrt{d\tan\theta (2\cos^2\theta + d\tan\theta +  d \sin2\theta)} \ ,
\end{equation}
They have the following  asymptotic expressions
\begin{equation}
\varepsilon_1 \approx H + O(d^2) \ , \qquad
\varepsilon_2 \approx H - 8J'S  + O(d^2) \ .
\end{equation}
Thus, the splitting between the two resonance branches in $\rm Cs_2CuCl_4$/$\rm Cs_2CuBr_4$ is again approximately given by
$$
\Delta\varepsilon \approx 4J' + O(d^2) \ .
$$

We emphasize again that despite a small contribution into $\Delta\varepsilon$,
a finite value of the staggered component of the DM vector $D_b'$ is essential
for observing the weak secondary mode. Its intensity vanishes as $D_b'\to 0$.

\section{Saturation field}

In the presence of noncollinear DM vectors (\ref{Hspin}), the calculation of
the saturation field for a general direction of the applied field becomes
rather cumbersome. Therefore, we restrict ourselves to the experimentally
relevant case  ${\bf H}\parallel {\bf b}$. This field orientation
is also the simplest one from the theoretical point of view
as  spins become parallel to the field at $H>H_s$, see Secs.~II and III.
For such a collinear spin arrangement it is easy to check that
the transverse DM terms with $D_a$ ($D_a'$) and  $D_c$ ($D_c'$)
contribute only cubic bosonic terms after the Holstein-Primakoff transformation
and, therefore, do not affect magnon energies in the harmonic approximation.
Thus, we need to take into account only the DM term with the longitudinal $D_b'$
component (\ref{HDMb}).

Leaving behind the standard steps we note that the
representation (\ref{HDMb}) implies that the $J'$ and $D_b'$ bonds
simply sum up in the saturated phase.
This leads to the replacement $J' \to J'+iD_b'$ in the hopping terms in the bosonic
Hamiltonian. As a result the magnon energy in the saturated phase is given by
a simple generalization of the standard equation
\begin{eqnarray}
& & \varepsilon_{\bf k} = H- 4(J+J')S + 4JS \cos^2\frac{k_x}{2}
- 4S\sqrt{J'^2+D_b'^2} \cos\frac{k_x}{2}\cos\frac{k_y\sqrt{3}}{2}\,.
\label{EkB}
\end{eqnarray}
Minimization of the above expression with respect to $\bf k$ yields
$k_y=0$ and $\cos k_x/2 = \sqrt{J'^2+D_b'^2}/2J$. As a result the saturation field
is given by
\begin{equation}
H_s = 4(J+J')S + \frac{J'^2 +D_b'^2}{J}\, S \,,
\end{equation}
which for $S=1/2$ yields the expression (8) from the main text.


\begin{thebibliography}{10}
\bibitem{acc} The largest error source in determing the exchange coupling parameters are the accuracy of the fit and the assumption of the DM contribution (which is less than 10-15 $\%$ of $J$). We estimate the accuracy of $J$ and $J'$ better than $\pm 5\%$.
\end{thebibliography}
\end{document}